\documentstyle[twocolumn,prl,aps]{revtex}
\input epsf
\begin{document} 
\title{
Symmetry of the Atomic Electron Density in Hartree, Hartree-Fock, 
and Density Functional Theory 
}
\author{
H.A.~Fertig$^{1,2}$ and W. Kohn$^1$
}
\address{
1. Department of Physics, University of California,
Santa Barbara, CA 93106
}
\address{
2. Department of Physics and Astronomy, University of Kentucky,
Lexington, KY 40506-0055
}
\address{\mbox{ }}
\address{\parbox{14.5cm}{\rm \mbox{ }\mbox{ }
The density of an atom in a state of well-defined total angular momentum
has a specific
finite spherical harmonic content, without and with
interactions.  {\it Approximate} single-particle
schemes, such as the Hartree,
Hartree-Fock, and Local Density Approximations, generally violate
this feature.  We analyze, by means of perturbation
theory, the degree of this violation
and show that it is small.  The correct symmetry 
of the density can be assured
by 
a constrained-search formulation without significantly altering
the calculated energies.  We compare our procedure to the 
(different) common
practice of spherically averaging the self-consistent potential.
Kohn-Sham density functional theory with the {\it exact} exchange-correlation
potential has the correct finite spherical harmonic
content in its density; but the corresponding exact single
particle potential and wavefunctions contain an infinite number
of spherical harmonics.
}}
\address{\mbox{ }}
\address{\mbox{ }}

\maketitle

\section{Introduction}

Single-particle descriptions of electronic states and densities 
in atoms date back
to their earliest models.  Most of them involve the
motion of individual electrons in some effective potential due
the nucleus and the other electrons; Bohr's early analysis
of some atomic spectra involved this idea\cite{bohr}.  With the advent
of wave mechanics, the idea took on the
form of solving single-particle Schroedinger equations with
this effective potential.
Prominent examples are the Hartree and
Hartree-Fock (HF) approximations\cite{hartree,comhar,hfa}, and 
density functional theory (DFT)\cite{kohn}.  Of these, only 
DFT provides in principle an exact description of electron
densities with their proper truncated spherical
harmonic content.  In practice, one is forced to 
adopt an approximate form for the exchange-correlation potential,
such as the local density approximation (LDA). 
In carrying out such calculations, one 
computes the electronic states and effective potential iteratively,
yielding a self-consistent potential and density.

The spherical symmetry of the nuclear potential yields
states of well-defined angular momentum.  However, except for
$S$ states, the resulting electron densities are generally
{\it not} spherically symmetric.  
($S$-states and their spherical densities present no
problem and will not be further considered.)
As we shall see below, in states of well-defined angular
momentum quantum numbers $L$ and $L_z$,
the exact density $n(\vec{r})$  
may be decomposed in the {\it finite} series
\begin{equation}
n(\vec{r})=\sum_{l=0}^L n_{2l}(r)Y_{2l}^0(\vec{\Omega}),
\label{den1}
\end{equation}
where the $n_{2l}(r)$ are radial functions and $Y_{2l}^0(\vec{\Omega})$
are spherical harmonics.

The self-consistent densities obtained via the  
approximation schemes  
described above do not have this form. 
This form {\it can} be and often is
obtained by introduction of a further approximation
which, however, violates self-consistency: the
effective potential may be spherically averaged, yielding
single-particle states with good angular momentum quantum
numbers and a resulting density of the form in Eq. \ref{den1}.
Such spherical averaging is of practical utility as
it greatly reduces the numerical effort involved in carrying
out the approximation schemes \cite{harrison}.  
Nevertheless, the Hartree, HF,
and LDA may all be expressed in terms
of variational principles, implying that the use of spherical
averaging leads to an overestimation of atomic energy levels.
To our knowledge, the quantitative effect of this has only
been checked in a small number of cases\cite{hartree2,janak}. 
The effect is thought to be small since the
resulting energies for many atoms are in quite good
agreement with experiment\cite{condon}.

In this paper, we will examine this, to our knowledge
largely unexplored, issue in some detail.  Using a perturbative
approach, we will demonstrate that the inappropriate spherical harmonic
components appearing in the self-consistent density 
(without spherical averaging) are
generally quite small.  
(An {\it exact} DFT calculation would, of course, yield the exact
density with the correct spherical harmonic content.)
We then
develop a constrained search principle to modify the variational
principles involved in the Hartree, HF, and DFT approximations
to guarantee that the resulting
density has the correct form of Eq. \ref{den1}.  We show 
in the context of the Hartree approximation that
this approach generates energies that
are only slightly higher than those from the unconstrained approximation.  

It is interesting to consider in more detail 
the implications of Eq. \ref{den1}
for {\it exact} DFT.  Being exact\cite{kohn}, it
is unnecessary to introduce constaints to
guarantee this ``symmetry'' of the density.  
What is the angular symmetry of
the exact
effective single particle potential entering the Kohn-Sham
equations which guarantees that the density will have this form?
A natural, but incorrect, guess would be that the Hartree
and exchange-correlation potentials together sum to a
potential that is spherically symmetric.  In fact this is
generally not true: the effective single particle potential
contains spherical harmonic components of {\it all} even orders.
(A concrete example of this is presented
in Appendix \ref{appendix:hooke}.)  Indeed, 
it has been shown\cite{goerling,the2} that
a unique, spherically symmetric single-particle potential $v_0(r)$
may be chosen to match the spherical component $n_0(r)$
of the density; it is possible to formulate an alternative
to the Hohenberg-Kohn theorem based solely on
$n_0(r)$ \cite{goerling}.  However,
the resulting potential yields incorrect higher order
components of the density $n_{2l}(r)$  ($l > 0$).
Thus, there is a kind of complementarity: if one insists that the density
have the correct truncated spherical
harmonic content for an interacting state of well-defined
angular momentum, the effective single particle potential
will not have spherical symmetry
and the single determinant model wavefunction will {\it not}
have good angular momentum quantum numbers.
By contrast, if one insists that the
single-particle potential be spherically symmetric,
the correct spherical harmonic content of the density 
of the interacting state cannot not be reproduced.

The remainder of this paper is organized as follows.  In Section II,
we first give the proof of Eq. \ref{den1}.   By
a perturbative approach to the Hartree approximation,
we demonstrate that the deviation of the density from its
appropriate symmetry is actually quite small in a specific
example (a Helium triplet state), and comment on
related results for the HF approximation and LDA.
In Section III, we formulate
a constrained-search approach to single-particle 
approximations for the density, which we apply to
the Hartree approximation and the LDA.  
We summarize our results in Section IV.
Finally, 
two Appendices are included.  
In Appendix \ref{appendix:hooke} we discuss
a two-electron harmonic atom 
with interactions (``harmonium'') and show
explicitly that the density of
its lowest triplet state cannot
be reproduced by a non-interacting system in
a spherically symmetric effective
single-particle potential.  
Appendix \ref{calculation}
contains some details of the numerical calculations.

\section{Spherical Harmonic Content of the Density}

\subsection{Finite Spherical Harmonic Content of the Density}

We begin by proving
Eq. \ref{den1} of the Introduction
by a standard application
of the Wigner-Eckart theorem.
Consider an atom in a state $|L,M>$ with total orbital angular momentum $L$ and
total azimuthal angular momentum $L_z=M$ along the
$z$-direction.  
We ignore spin-orbit coupling, so that the orbital and
spin states of the atom may be specified separately. 
We are 
interested in the expectation value of the
operator $n(\vec{r})=\sum_i \delta(\vec{r}-\vec{r}_i)$,
where $\vec{r}_i$ denotes the position of the $ith$ electron.
A useful decomposition of the delta function in this context
is\cite{arfken}
$$
\delta(\vec{r}-\vec{r}_i) = {1 \over {r^2}}
\delta(r-r_i)
\sum_{l=0}^{\infty} \sum_{m=-l}^{l}
Y_l^{-m}(\vec{\Omega}) Y_l^{m}(\vec{\Omega}_i)
$$
where 
$\vec{\Omega}$ represents an angular direction in spherical
coordinates.  The set
$\lbrace Y_l^m(\vec{\Omega}_i) ,~m=-l,-l+1,...l \rbrace$ constitutes an
irreducible tensor operator with respect to the angular
momentum operator $L$, and obeys the Wigner-Eckart theorem\cite{merzbacher}.
It follows that 
$\langle L,M | Y_l^m(\vec{\Omega}_i) | L,M \rangle$ is 
proportional to the Clebsch-Gordon coefficient
$\langle LlMm | LlMm \rangle,$ 
which vanishes
unless $m=0$, $0 \le l \le 2L$, and $l$ is even.  Substituting
the expansion for the delta function into the expectation
value of the density and using the above observation directly
yields Eq. \ref{den1}.

\subsection{Infinite Spherical Harmonic Content in the Hartree 
and Hartree-Fock Approximations}
\label{pert_har}

The decomposition of the physical
density of an atomic state in spherical harmonics consists of
a finite series.  However, the Hartree, HF, and approximate DFT
solutions do {\it not} produce densities with
this property.  
For example,
suppose  
we $could$ find a finite decomposition 
for the density in the Hartree approximation,
$$n^H(\vec{r})=\sum_{l=0~(l~even)}^{l_{max}} n_{l}(r) Y_{l}^0(\vec{\Omega}).$$
The effective single
particle potential contains a term of the form
$$\lambda e^2 \int d^3 r'
{{ n^H(\vec{r'})} \over {|\vec{r} -\vec{r}'|}}$$
where $\lambda$ $(0 \le \lambda \le 1)$ is a parameter
by which we may switch on the electron-electron
interaction, which will be useful below.
This term has a spherical
harmonic decomposition with maximum $l=l_{max}$.
The effective potential in the single particle Schroedinger
equation multiplies a wavefunction $\phi$;
for a single particle state of azimuthal quantum number $m=0$
this may be expanded as
\begin{equation}
\label{interaction}
\phi(\vec{r}) = \sum_{l'=0}^{l'_{max}}
 y_{l'}(r) Y_{l'}^0(\vec{\Omega}).
\end{equation}
When multiplied by the potential,
the resulting products of spherical harmonics 
may be expressed as linear combinations of single spherical
harmonics $Y_{l}^0$, with a maximum
non-vanishing contribution from $l=l_{max}+l'_{max}$.
The other terms in the Schroedinger
equation however contain spherical
harmonics of order $ l $ no greater than $ l'_{max}$.
Thus, the Schroedinger equation {\it cannot}
be solved by wavefunctions expressable in
a finite spherical harmonic expansion
(except for the trivial case of $L'=0$).

The density produced from these wavefunctions
in general has no finite spherical
harmonic expansion.  One way to demonstrate
this uses perturbation theory.  The solution
to the Hartree equations may be expressed as
a power series in $\lambda$;
terms of higher order 
involve increasingly larger orders
of spherical harmonics.  
When reorganized as a spherical harmonic expansion,
all orders will occur with each
coefficient a power series in $\lambda$.
It is not possible for
these coefficients to
vanish for arbitrary values of $\lambda$.

As a concrete example,
we analyze a two electron atom in a triplet spin state
with total angular momentum $L=1$, whose density is not spherically
symmetric.  
Using perturbation
theory in the electron-electron interaction,
we compute the density in
the Hartree approximation.  
The Hamiltonian for our system is 
\begin{equation}
H=\sum_{i=1,2} \bigl[
-{{1} \over {2m}} \nabla_i^2 - {{Ze^2} \over {r_i}} \bigr]
+\lambda e^2  {1 \over {|\vec{r}_1-\vec{r}_2|}},
\label{hamiltonian}
\end{equation}
where $Z$ is the nuclear charge.
The fully interacting system is given by $\lambda=1$, and
we will formally develop our perturbation theory in powers
of $\lambda$.  Alternatively, one may set $\lambda=1$ and
consider an expansion of energy and density in powers of
$1/Z$, which is equivalent to an expansion in $\lambda$.
Physically, one should thus think of the small $\lambda$ limit
as the state of a highly ionized atom of large $Z$.
The specific case we will focus on is $N=2$; thus
$\lambda=1$ and $Z=2$
describes the helium atom.

In the absence of interactions, the state of interest to us
involves one electron in a $1s$ state and one in a $2p$ state
which we take to be
in the $m=0$ state.  It is easy to see that the density of
this state satisfies Eq. \ref{den1}:
\begin{eqnarray}
\label{nonintden}
n^{(0)}(\vec{r}) &=& |\phi_0^{(0)}(\vec{r})|^2+
                     |\phi_1^{(0)}(\vec{r})|^2 \nonumber \\
                 &=& |R_{10}(r) Y_0^0(\vec{\Omega})|^2+
                     |R_{21}(r) Y_1^0(\vec{\Omega})|^2 \nonumber \\
                 &=& [c_{00}^0 R_{10}(r)^2 + c_{11}^0 R_{21}(r)^2]
                          Y_0^0(\vec{\Omega}) \nonumber \\
                  && \quad\quad  +c_{11}^2 R_{21}(r) Y_2^0(\vec{\Omega}). 
\end{eqnarray} 
In Eq. \ref{nonintden}, $\phi_0^{(0)}$ and $\phi_1^{(0)}$ are respectively
the $1s$ and $2p$ states, $R_{nl}(r)$ are hydrogenic radial
functions, with $n$ the principal quantum number and $l$ the
angular momentum; the coefficients $c_{jk}^i$ are defined
as
$$
c_{jk}^i = \int d \vec{\Omega} Y_i^0(\vec{\Omega})
                Y_j^0(\vec{\Omega}) Y_k^0(\vec{\Omega}),
$$
so that 
$Y_j^0(\vec{\Omega}) Y_k^0(\vec{\Omega})  \equiv \sum_i
c^i_{jk} Y_i^0(\vec{\Omega})$.  The coefficients $c_{jk}^i$ are
closely related to Gaunt coefficients commonly used in
atomic structure calculations\cite{condon}, and have properties
similar to Clebsch-Gordon coefficients; in particular, 
$c^i_{jk} = 0$ unless $i+j+k$ is even and $|j-k| \le i \le |j+k|$.
It is these two properties that guarantee the density of the
noninteracting state has the truncated form in Eq. \ref{den1}.  
The superscript,
$(0)$, in Eq. \ref{nonintden} denotes non-interacting quantities
($\lambda=0$).

The Hartree approximation amounts to self-consistently finding
two single-particle eigenstates $\phi_0$ and
$\phi_1$, of energies $\varepsilon_0,~\varepsilon_1$,
for non-interacting electrons moving in an effective
potential
\begin{equation}
V_{eff}^H =  - {{Ze^2} \over {r_i}} + 
\lambda e^2 \int d^3r' {{ n(\vec{r'})} \over {|\vec{r} -\vec{r}'|}}
\label{veff}
\end{equation}
where the density is 
$n(\vec{r}) = |\phi_0(\vec{r})|^2  + |\phi_1(\vec{r})|^2$.
To first order in perturbation theory, we may write that potential as
$$
V_{eff}^H =  - {{Ze^2} \over {r_i}} + 
\lambda e^2 \int d^3r' {{ n^{(0)}(\vec{r'})} \over {|\vec{r} -\vec{r}'|}} 
+ {\cal O} (\lambda^2) 
$$
$$
\equiv - {{Ze^2} \over {r_i}} + 
\lambda U^{(1)}(\vec{r}) + {\cal O} (\lambda^2).
$$
The Schroedinger equation arising in the Hartree approximation may be
solved within perturbation theory by expanding the effective
potential, eigenstates, and eigenenergies in powers of $\lambda$.
The first order correction to the eigenstates 
satisfies the inhomogeneous differential equation
\begin{equation}
[H_0 - \varepsilon_i^{(0)}]\phi_i^{(1)}(\vec{r})
=[\varepsilon^{(1)}_i - U^{(1)}] \phi_i^{(0)} (\vec{r}),
\label{wfeqn}
\end{equation}
where $i=0,1$, and
the first order correction to the energies are
$\varepsilon_i^{(1)} = \int d^3r
\phi^{(0)*}_i(\vec{r}) U^{(1)} (\vec{r}) \phi^{(0)}_i(\vec{r})$. 
Because the density $n^{(0)}(\vec{r})$ contains only
even spherical harmonics, so will the potential $U^{(1)}$.
It immediately follows that the the wavefunction corrections
$\phi_i^{(1)}$ will have the same parity as the states $\phi_i^{(0)}$
from which they descend.  Using the multiplicative properties
of the $Y_l^0$ gives
\begin{eqnarray}
\phi_0^{(1)}&=&y_0(r)Y_0^0(\vec{\Omega})+y_2(r)Y_2^0(\vec{\Omega}),
  \nonumber \\
\phi_1^{(1)}&=&y_1(r)Y_1^0(\vec{\Omega})+y_3(r)Y_3^0(\vec{\Omega}),
\label{y1}
\end{eqnarray}
where $y_i(r)$ are purely radial functions. 

Our development of the perturbation theory already illustrates one 
of the central points of this paper: we can see 
that the effective potential (which we have computed to first order
in $\lambda$) is not spherically symmetric and
the wavefunctions 
arising in the Hartree equation do not have well-defined angular
momentum.  
By expanding both sides of Eq. \ref{wfeqn} in spherical
harmonics and matching the coefficients for each $l$,
the equations for the radial functions may all be written in the
form
\begin{equation}
[h_0(l)-\varepsilon_l^{(0)}] y_l(r) = f_l(r)
\label{y2}
\end{equation}
where 
$$h_0(l)=-{1 \over {2m}}{1 \over {r^2}} {d \over {dr}}
r^2 {d \over {dr}} + {{l(l+1)} \over {2mr^2}} - {{Ze^2} \over {r}}$$
and $\varepsilon_l^{(0)} = \varepsilon_0^{(0)}$ for even $l$,
$\varepsilon_1^{(0)}$ for odd $l$.  The functions $f_l$
are easily computed, and the equations may be solved numerically.
This calculation will be presented in the next section.
Once the radial functions $y_l$ are obtained, the first
order correction to the density in the Hartree approximation
is found by adding the squared wavefunctions and collecting
terms of order $\lambda$.  The resulting density may be
written in the form
$$
n^{(1)}(\vec{r}) = n^{(1)}_0({r})Y_0^0(\vec{\Omega})
                 + n^{(1)}_2({r})Y_2^0(\vec{\Omega})
                 + n^{(1)}_4({r})Y_4^0(\vec{\Omega})
$$
with
\begin{eqnarray}
n^{(1)}_0(r) &=& 2c^0_{00}R_{10}(r)y_0(r)+2c^0_{11}R_{21}(r)y_1(r) \nonumber \\
n^{(1)}_2(r) &=& 2c^2_{02}R_{10}(r)y_2(r)+2c^2_{11}R_{21}(r)y_1(r) 
                 +2c^2_{13}R_{21}(r)y_3(r)\nonumber \\
n^{(1)}_4(r) &=& 2c^4_{13}R_{21}(r)y_3(r) 
\label{den2}
\end{eqnarray}

Note that to this order in $\lambda$, only one ``offending'' spherical
harmonic, $Y_4^0(\vec{\Omega})$, 
appears.  However, all even spherical harmonics would appear in
higher orders in perturbation theory.
Figure 1 illustrates the radial functions $n^{(1)}_l(r)$ for the
present model problem, as well as the analogous zeroth order densities 
$n^{(0)}_l(r)$ appearing in the spherical harmonic decomposition
of the density for the non-interacting problem (cf. Eq. \ref{nonintden}).
Note that the magnitude of 
the offending spherical harmonic component
is quite small (ratio of maximum contribution to root
mean square density 3.86 $\times 10^{-3}$), and that
the densities $n^{(1)}_l(r)$
decrease very rapidly with increasing $l$.  The reason for this
is that at zeroth order, the density (of the non-interacting system)
varies rather slowly as a function of the angular variable.  When
interactions are introduced, such a slowly varying potential has
only a small amplitude for scattering electrons into high angular
momentum states; the resulting density thus only has a small 
component of large $l$ spherical harmonics.  It is clear that
this property is true at all orders in perturbation theory:
the effective potential entering at any order will always have
a much larger $Y_0^0$ component than any other, leading to only
small admixtures of high angular momenta in the wavefunctions.
Finally, although we have illustrated this property in the specific context
of a helium triplet $P$ state, it should be quite
general for atoms.  
Indeed, our model problem is in some sense
a ``worst-case'' example; for larger atoms, particularly ones
with many closed shells, the predominant spherical components of the 
density will be even larger.
This helps to explain
the success of using spherically averaged effective potentials
in the Hartree approximation\cite{hartree2}.

Spherical averaging is also a common practice in applying the Hartree-Fock 
approximation\cite{condon}, and it is therefore of interest to assess
the extent to which Eq. \ref{den1} will be 
violated without such averaging.  
We again proceed perturbatively.  In addition to the direct
potential $U^{(1)}(\vec{r})$, there is now a non-local exchange potential,
to first order in perturbation theory, and the
corrections to the wavefunctions take the form 
\begin{eqnarray}
\phi_0^{(1),HF}&=&y_0^{HF}(r)Y_0^0(\vec{\Omega})+y_2^{HF}(r)Y_2^0(\vec{\Omega}),
  \nonumber \\
\phi_1^{(1),HF}&=&y_1^{HF}(r)Y_1^0(\vec{\Omega}).
\label{y1hf}
\end{eqnarray}
There is no $Y_3^0$ term in the wavefunctions because
there is a precise cancellation between the direct and exchange
terms.  The resulting
density, remarkably, has precisely the right form -- Eq. \ref{den1} --
to first order, unlike in the Hartree approximation.
In fact,
one may demonstrate that the solution
to the HF approximation reproduces the correction to the
density {\it exactly} to
first order in the electron-electron interaction. 

Unfortunately, this good property of the solutions to the HF equations
is limited to first order in $\lambda$.  This is most easily seen in
the context of our model calculation for the He $P$ state.  The
presence of a $Y_2^0$ component to the density ensures that the
effective potential seen by either electron has a similar
component.  For the $\phi_0$ state, this leads
to a contribution proportional to $Y_2^0(\vec{\Omega})$,
as in Eq. \ref{y1hf}.  At second order in $\lambda$, this necessarily
produces a component in the density proportional
to $Y_4^0(\vec{\Omega})$.  
However, the fact
that the HF density has the correct form to order $\lambda$ indicates
that the magnitude of the violation will be
even smaller than that found in the Hartree
approximation. 

\subsection{Harmonic Content of the Density in
Density Functional Theory and Local Density Approximation}

The violations of Eq. \ref{den1} found in the Hartree and
HF theories do {\it not} occur for the {\it exact}
DFT, which, by construction, produces exact
densities.  In practice, one must always introduce approximations
for the exchange-correlation energy and potential.
To illustrate the point,
we consider a perturbative application of the local density
approximation (LDA) to our helium $P$ state example.  

The formalism
closely parallels our perturbative approach to the Hartree
approximation.  In LDA, we need to solve self-consistently
a Schroedinger's equation with an effective potential
given by $V_{eff}^H+V_{xc}^{LDA}$, where $V_{eff}^H$ is given
by Eq. \ref{veff} and $V_{xc}^{LDA}=\delta E_{xc}^{LDA}[n(\vec{r})]/
\delta n(\vec{r})$.  For the purpose of this illustration,
we neglect the correlation contribution and take 
$V_{xc}^{LDA}(\vec{r}) \approx V_x(\vec{r})
=-\bigl[{6 \over {\pi}}n(\vec{r})\bigl]^{1/3}$
\cite{parr}.
To first order in
$\lambda$, it is sufficient to replace $n$ in $V_{xc}^{LDA}$
with $n^{(0)}$.  Because $V_{xc}^{LDA}$ is not an analytic
function of the density, it is important to recognize that
when $V_{xc}^{LDA}$ is expanded in terms of spherical harmonics
$Y_l^0(\vec{\Omega})$, 
the resulting series will involve all even values
of $l$.  This contrasts with the Hartree contribution,
which at this order contained only $l=0$ and $l=2$
components.  In some sense this suggests LDA will
lead to stronger violations of Eq. \ref{den1} than
we encountered in the Hartree approximation.  However,
the effective potential we construct at first order
in $\lambda$ is a slowly varying function of $\vec{\Omega}$,
so that contributions from large values of $l$ to the
wavefunctions are still quite small.

When expanded in spherical harmonics, the corrections to the wavefunctions
have a form very similar to Eq. \ref{y1}, except 
$\phi_0^{(1)}$ will now contain {\it all} even spherical
harmonics, and $\phi_1^{(1)}$ will contain all odd ones.  Writing 
$\phi_i^{(1)}=\sum_l y_{2l+i}Y_{2l+i}^0(\vec{\omega})$,
the equations satisfied by the $y_{2l+i}$'s are identical
in form to Eq. \ref{y2}, with a modified form for
the inhomogeneous functions $f_l$.   Once the
radial functions have been obtained, the first order
correction to the density in LDA is given by
$n^{(1)}(\vec{r}) = \sum_l n_{2l}^{(1)}(r) Y_{2l}^0(\vec{\Omega})$
with
$$
n_{2l}^{(1)}(r)=2c_{0,2l}^{2l}R_{10}(r) y_{2l}(r)
+2c_{1,2l-1}^{2l}R_{21}(r) y_{2l-1}(r)
$$
$$
\quad\quad\quad
+2c_{1,2l+1}^{2l}R_{21}(r) y_{2l+1}(r).
$$
In practice, the expansion of the
density falls off so rapidly (see Fig. 3) with $l$ that
only the lowest few functions $y_{n}$ need to be computed.

\section{Restoring the Symmetry of the Density} 

\subsection{Constrained Search Formulation}

The calculations in the above sections to some extent explain
why spherical averaging is successful in the Hartree and HF approximations,
and in the LDA, when applied to atoms.
Nevertheless, the averaging is basically {\it ad hoc},
and lacks a clear justification.  From a formal point
of view, spherical averaging has a dissatisfying
aspect: the minimization principles that are
used to derive the three approximations are abandoned
when it is introduced.  Formally, a more consistent approach
-- especially for DFT --
is to modify, or more precisely, {\it constrain} the
wavefunctions searched in the minimizations
in such a way that Eq. 1 is guaranteed.  This has the
advantage that the energies of the atomic states found will be
lower than those found by spherical averaging.  In practice,
however, the energy lowering turns out to be quite small.
Nevertheless, it is useful to explore constrained search
methods for preserving symmetry properties of the density
because such violations are known to occur in other
symmetry properties -- particularly those involving spin\cite{davidson} --
and may be responsible for more serious errors that arise
in molecular calculations. 
The present formalism is a first example
of how to consistently impose symmetry on an approximate
single-particle scheme.

As stated above, the Hartree, HF, and LDA equations are
derived from minimization principles.  In the Hartree
approach, the energy functional is
\begin{equation}
E[\Psi]=
<\Psi|H_0|\Psi> + {{e^2} \over 2}\int d^3r d^3r' 
{{n(\vec{r})n(\vec{r}')}
\over {|\vec{r}-\vec{r}'|}}.
\label{efun1}
\end{equation}
Here, $|\Psi>$ is a normalized wavefunction, 
$H_0$ is the non-interacting electron Hamiltonian,
and $n(\vec{r})$ is the expectation value of
the density in the state $|\Psi>$.
To generate the Hartree equations, one
minimizes $E[\Psi]$ among
orthonormal product wavefunctions\cite{hartree2}.  
In density functional theory, one adds an appropriate
exchange-correlation energy $E_{xc}$ to the expression
\ref{efun1},
and then searches for the minimum of the resulting 
energy\cite{parr,gross}.
(This presumes the density is non-interacting $v$-representable,
which we will assume for the states of interest.)
For the exact exchange-correlation energy the resulting
density satisfies Eq. \ref{den1}.  Of course, the
exact exchange-correlation energy is unknown, and, 
in practice, one is forced to adopt approximations\cite{parr}.  

To constrain these searches to the subspace of states having
a density of the form of Eq. \ref{den1}, we introduce a set
of $r$-dependent
Lagrange multipliers $\Lambda_{2l}(r)$.  The constraints that must be
enforced are
\begin{equation}
\int d \vec{\Omega} n(\vec{r}) Y_{2l}^0(\vec{\Omega}) = 0, \quad 2l > 2L,
\label{constraint}
\end{equation}
where is $L$ is the angular momentum of the
state of interest.
The function we need to minimize is
\begin{equation}
E+E_{xc} + \int d^3 r V_R(\vec{r}) n(\vec{r}),
\label{constrained_en}
\end{equation}
where
\begin{equation} 
V_R(\vec{r}) \equiv \sum_{l>2L} \Lambda_{2l}(r) Y_{2l}^0(\vec{\Omega}).
\end{equation}

After minimization of Eq. \ref{constrained_en}, we arrive
at a single-particle Schroedinger equation 
\begin{equation}
\bigl[
-{{1} \over {2m}} \nabla^2  + v_s(\vec{r}) \bigr ]
\phi_i(\vec{r}) = \varepsilon_i \phi_i(\vec{r}).
\label{schro}
\end{equation}
For density functional theory, 
\begin{equation}
v_s(\vec{r}) = - {{Ze^2} \over {r}} + 
e^2 \int d^3 r' {{n(\vec{r})} \over {|\vec{r}-\vec{r}'|}}
+V_{xc}([n(\vec{r})],\vec{r}) + V_R(\vec{r}),
\label{effpot}
\end{equation}
where $V_{xc}$ is the exchange-correlation potential.
For an $N$-electron atom, filling the lowest $N$ eigenstates
of Eq. \ref{schro} leads to the density used in Eq. \ref{effpot},
so these equations must be solved self-consistently.  
The Lagrange parameters $\Lambda_{2l}(r)$
of Eq. \ref{constrained_en} must be chosen to satisfy
Eq. \ref{constraint}. 

One natural, but incorrect, guess would be that $V_R$
simply removes the high spherical harmonics present in the other
terms entering $v_s$, rendering a spherically symmetric
single particle potential.  This is not possible except for the
trivial case of $S$ states.  For example,
in our model calculation of the helium $P$ state, the lowest
spherical harmonic component present in $V_R$ is $l=4$, which
cannot remove the $l=2$ component coming from the Hartree term.
In fact,
the single particle potential in general contains $all$ orders
of spherical harmonics.  With the
exact form of $V_{xc}$,  $V_R=0$, but $V_{xc}$ 
itelf contains an infinite number of spherical harmonics.
This is demonstrated in a specific soluble model 
(Appendix \ref{appendix:hooke}):
two fermions in a harmonic trap, interacting via a repulsive quadratic
potential.  The exact eigenfunctions of this system may be written
down explicitly, and in the Appendix we show (a) that
the density of a $P$ state cannot be produced by a non-interacting
electron system in any spherically symmetric potential,
and (b) that the unique single particle potential 
reproducing this density 
(cf. the Hohenberg-Kohn theorem\cite{kohn})
in a non-interacting system contains all even orders of spherical
harmonics.

\subsection{Perturbative Implementation}

Our proposed solution to the problem of producing densities that
have an appropriate form for orbital angular momentum eigenstates
thus reduces to finding a self-consistent solution to Eqs. \ref{constraint},
\ref{schro}, and \ref{effpot}.  
The following is a practical procedure: (1) Obtain the
self-consistent Kohn-Sham solution with the given $E_{xc}[n]$
and the self-consistent total potential $v_s(\vec{r})$.
We expect that this will violate {\it weakly}
the constraints (\ref{constraint}), with the offending
density components $n_{2L+2}(r),~n_{2L+4}(r),\dots$
being small.  (2) The restoring potential, 
$V_{R,2L+2}(r),~V_{R,2L+4}(r),\dots$ is determined
by solving the equation
\begin{equation}
\left( 
\begin{array}{c}
-n_{2L+2}^0(r) \\ -n_{2L+4}^0(r) \\ -n_{2L+6}^0(r) \\ \vdots
\end{array}
\right)
=\int_0^{\infty} dr' r'^2 {\bf K}^{FF}([v_s(0)];r,r')
\left(
\begin{array}{c}
V_{R,2L+2}^0(r') \\ V_{R,2L+4}^0(r') \\ V_{R,2L+6}^0(r') \\ \vdots
\end{array}
\right)
\label{matrix}
\end{equation}
where $K^{FF}$ is the submatrix $(l,l'>2L)$ of the linear
density response function $K_{l,l'}(r,r')$
corresponding to $v_s(\vec{r})$.  (3) The new
wavefunction and density, satisfying the constraints
(\ref{constraint}) are determined from $v_s(\vec{r})+V_R(\vec{r})$.

Equivalently, this process can be carried out in
terms of wavefunctions (see Appendix \ref{calculation}). 
Perturbative energies for our helium $P$ state example
using the constrained Hartree approximation
and LDA are presented in Table I along with comparable results
for unconstrained and spherically averaged approaches.

\section{conclusion}

This paper deals with the angular dependence of the electron
density $n(\vec{r})$ of an atom in a state of finite
angular momentum $L$, both in the exact physical state
and in various single particle descriptions.

The spherical harmonic content of the physical density, as
a direct consequence of the Wigner-Eckart theorem, is 
limited to even values of $l \le 2L$.  However in the
Hartree, Hartree-Fock, and the various approximate forms of Kohn-Sham theory,
the spherical harmonic content of the density involves {\it all}
$l$-values (although components with $l >2L$ are small.)
The exact Kohn-Sham effective single particle potential
by definition reproduces  the $l$-limited physical density;
on the other hand, the potential involves all $l$-values.
(This is documented for the case of an exactly soluble
model of an atom with interacting electrons, the
``harmonium'' atom.)

We show how the requirement, $l \le 2L$, can be restored
by a constrained search procedure using Lagrange
parameter functions.  Various numerical illustrations
are presented.

Somewhat analogous symmetry violations are known to arise
in connection with the electronic spin quantum numbers.
They may be susceptible to similar analysis and
symmetry restoration.

This work was supported by the NSF through Grant Nos. 
DMR 960452, DMR 9870681, and DMR 9976457, and by the
Research Corporation. 
We thank Drs. D. Claugherty and Y. Meir for discussions
of the finite spherical harmonic content of the physical density.

\vfill\eject
                    
\appendix

\section{Density for a Harmonic Atom}
\label{appendix:hooke}
 
In this Appendix, we present a calculation of a $P$ state for two
spinless fermions trapped in a quadratic potential, interacting
via a repulsive quadratic potential.    We call such a
harmonic atom ``harmonium''.  This is not intended as a
realistic model of a physical atom.  But it shares with
physical atoms their symmetry properties and allows
analytic calculations of wavefunctions and densities.
Our main goals in this calculation
are to demonstrate that ({\it i}) the exact density in this interacting state
cannot be reproduced by non-interacting fermions in any spherically
symmetric potential, and ({\it ii}) that the single particle potential that
{\it does} reproduce the density contains spherical harmonics
of all even orders.  

Our model Hamiltonian is
\begin{equation}
\label{hooke_ham}
H = -{1 \over {2m}}[\nabla_1^2+\nabla_2^2] + {1 \over 2} m \omega_0^2
[r^2_1+r_2^2] -   {1 \over 2} m \omega_1^2 |\vec{r}_1-\vec{r}_2|^2.
\end{equation}
Defining center of mass and relative coordinates $\vec{r}_{cm}=
(\vec{r}_1 + \vec{r}_2)/2,~\vec{r}=(\vec{r}_1 - \vec{r}_2)/2$,
this may be rewritten as a sum of commuting Hamiltonians, one
for the center of mass coordinate ($H_{CM}$) and one for
relative coordinates ($H_R$),
\begin{eqnarray}
\label{hooke_ham2}
H_{CM}&=& -{1 \over {2\mu}}\nabla_{cm}^2 + {1 \over 2} \mu \omega_0^2 r_{cm}^2
\nonumber \\
H_{R}&=& -{1 \over {2\mu}}\nabla^2 + {1 \over 2} \mu \omega_R^2 r_{cm}^2,
\end{eqnarray}
where $\mu=2m$, $\omega_R^2=\omega_0^2 - {1 \over 2} \omega_1^2$.
The total angular momentum operator
may be written in the form $\vec{L}=\vec{L}_{CM}+\vec{L}_{R}$,
the sum of angular momentum operators for the center of mass
and relative coordinates.  Using the composition rules for
angular momenta\cite{merzbacher}, it is easy
to see that the $P$ state of lowest energy  is formed by
putting the center of mass degree of freedom
in an $s$ state and the relative degree of freedom in a $p$ state.
Using
the explicit forms for harmonic oscillator states, the wavefunction
is
\begin{eqnarray}
\Psi &=& \biggl\lbrace \bigl[ {1 \over {\pi l_{CM}^2}} \bigr]^{3/4}
\exp\bigl[{-{{r_{cm}^2} \over {2 l_{CM}^2}}} \bigr] \biggr\rbrace
\nonumber
\\
&& \times
\biggl\lbrace {1 \over {\sqrt{2}}} \bigl[ {1 \over {\pi l^2}} \bigr]^{3/4}
H_1\bigl({z \over l}\bigr)
\exp \bigl[{-{{r^2} \over {2 l^2}}}\bigr] \biggr\rbrace,
\label{hooke_wf}
\end{eqnarray}
where $H_1(x)=2x$ is a Hermite polynomial, $l^2 = (\mu\omega_R)^{-1}$,
and $l_{CM}^2 = (\mu\omega_0)^{-1}.$  
The density of this state is
\begin{eqnarray}
n(\vec{r})&=&[A+Bz^2]e^{-r^2/L^2} \nonumber \\
&=&\bigl\lbrace [A + {1 \over 3}Br^2] + {B \over 3} 
\sqrt{{16\pi} \over 5}r^2 Y_2^0(\vec{\Omega}) \bigr\rbrace e^{-r^2/L^2}
\label{hooke_den}
\end{eqnarray}
with 
\begin{eqnarray}
A&=&4\pi^{3/2}\xi^5C^2 \nonumber \\
B&=&(4\pi)^{3/2} {{\xi^3 l^4} \over {L^4}} C^2.
\label{AB}
\end{eqnarray}
The length scales appearing in the above two equations are given
by $L^{2} = l^2+l_{CM}^2$ and $\xi^{-2}=l_{CM}^{-2} + l^{-2}$, and
$C=[ \pi l_{CM} l ]^{-3/2} [\sqrt{2} l]^{-1}$.
Note that Eq. \ref{hooke_den} has the form required by
Eq. \ref{den1}.  

We now demonstrate that Eq. \ref{hooke_den} is not derivable from a
system of {\it non-interacting} fermions in a spherically symmetric
external potential.  To show this, suppose the density {\it was}
derivable from such a potential.  Then the two occupied single
particle states would necessarily have the form $\phi_0(\vec{r})=
S(r)Y_0^0(\vec{\Omega})$, $\phi_1(\vec{r})=P(r)Y_1^0(\vec{\Omega})$.
The sum of the squares of these gives the density; matching
this to Eq. \ref{hooke_den} gives explicit expressions for $S(r)$ and
$P(r)$, 
\begin{eqnarray}
S(r) &=& \sqrt{4\pi A} e^{-r^2/2l^2} \nonumber \\
P(r) &=& \bigl[ {B \over {3 c_{11}^2}} \sqrt{{16\pi} \over 5} \bigr]^{1/2}
r e^{-r^2/2l^2}.
\label{SP}
\end{eqnarray}
It is interesting to notice that the functional forms of $S$ and
$P$ are perfectly compatible with a state of non-interacting
fermions in a harmonic trapping potential.  However, the
{\it normalizations} of $S$ and $P$ are not correct.  For example,
$S(r)$ is properly normalized if and only if $A=(\pi L^2)^{3/2}.$
An examination of the explicit expression for $A$, Eq. \ref{AB},
reveals that this is the case only if $\omega_1=0$;
i.e., the repulsion vanishes.
Thus, for non-interacting fermions, {\it no}
spherical potential will
reproduce the interacting density.

On the other hand, a non-interacting potential 
that is not spherically symmetric can be found to make
the density of two non-interacting fermions take the form
of Eq. \ref{hooke_den}.  We again
demonstrate this by using perturbation theory.  Because the 
density is axially symmetric -- i.e., it may be written
as a function of $r$ and $z$ -- we look for an effective
potential which is also axially symmetric.  In the body of this
work we have essentially expressed the $z$-dependence of
densities and potentials in terms of spherical
harmonics.  However, in this Appendix because harmonic oscillator
wavefunctions have a number of useful algebraic properties,
we expand
instead in powers of $z$.  

The form of Eq. \ref{hooke_den} suggests that 
the effective single-particle potential
that reproduces the density has the form
$$
V_{eff}(\vec{r}) = {1 \over 2} m \omega_{eff}^2 r^2 + \delta V(z),
$$
where $\omega_{eff} = 1/\mu L^2$.  We will perform our perturbation
theory around a $V_{eff}$ with $\delta V=0$; this is slightly
different than working around the non-interacting state, as
the length scale $L$ is modified by interactions.  Nevertheless,
it is clear that $\delta V$ must be small if the interaction
strength is weak.  For this form of the potential, it is also
clear that the two single-particle states must have the form
\begin{equation}
\phi_i(\vec{r}) = \psi_0(x)\psi_0(y)\chi_i(z)
\label{phi_i}
\end{equation}
with $i=0,1$, and $\psi_0$ the ground state of a one-dimensional
harmonic oscillator with frequency $\omega_{eff}$, and for 
$\delta V=0$, $\chi_0(z) \equiv \psi_0(z)$ may be taken as a harmonic
oscillator ground state and $\chi_1(z) \equiv \psi_1(z)$ 
as the first excited state.  If the electrons were non-interacting,
we would necessarily have in Eq. \ref{hooke_den}
\begin{eqnarray}
A=A_0&\equiv&\bigl[ {1 \over {\pi L^2}} \bigr] ^{3/2} \nonumber \\
B=B_0&\equiv&\bigl[ {1 \over {\pi L^2}} \bigr] ^{3/2} {2 \over {L^2}}.
\label{A0B0}
\end{eqnarray}
It is convenient to parameterize the effect of interactions in terms
of the deviations of $A$ and $B$ from these values,
$$
\delta n(\vec{r}) \equiv [\delta A + \delta Bz^2] e^{-r^2/L^2}
$$
with $\delta A \equiv A-A_0$, $\delta B =B-B_0$.  Denoting the
first order corrections to $\chi_i(z)$ as $\delta \chi_i(z)$,
it is easy to demonstrate to lowest order in perturbation
theory that
\begin{equation}
[\delta \tilde{A} + \delta \tilde{B} z^2 ] e^{-z^2/2L^2}
=\delta \chi_0(z) + \sqrt{2} {z \over L} \delta\chi_1(z),
\label{eq:a}
\end{equation}
where $(\delta \tilde{A}, \delta \tilde{B}) = 
(\pi L^2)^{5/4}(\delta A,\delta B)/2$. With some algebra Eq. \ref{eq:a}
may be written as
\begin{equation}
\delta \chi_0(z) + \sqrt{2} {z \over L} \delta\chi_1(z)
= \beta \psi_2(z)
\label{eq:b}
\end{equation}
where $\beta = {{2^{3/2}} \over 6}(\pi L^2)^{1/4} [\delta \tilde{B} L^2
- \delta \tilde{A} ]$, and $\psi_2(z)$ is the $n=2$ harmonic
oscillator state.

We now expand
$\delta \chi_0(z),~\delta \chi_1(z)$ in harmonic oscillator states:
\begin{eqnarray}
\delta \chi_0(z) &=& \sum_{n=0}^{\infty} c_{2n} \psi_{2n}(z) \nonumber \\
\delta \chi_1(z) &=& \sum_{n=0}^{\infty} c_{2n+1} \psi_{2n+1}(z).
\label{chi_exp}
\end{eqnarray}
The coefficients $c_n$ obey the recursion relation
\begin{equation}
\sqrt{n+1} c_{n+1} = \beta \delta_{n,2} -c_n - \sqrt{n} c_{n-1}
\label{recrel}
\end{equation}
for $n \ge 2$.  In addition, $c_0,~c_1=0$
since $\delta \chi_0(z),~\delta \chi_1(z)$
must be
orthogonal to $\chi_0,~\chi_1$.

Perturbation theory defines $\delta \chi_0,~ \delta \chi_1(z)$ in
terms of $\delta V$.  A useful expansion for $\delta V$ is
\begin{equation}
\delta V(z) = \sum_{n~{\rm even}}  {{v_n} \over {2^{n/2} \sqrt{n!}}}
H_n({z \over L}).
\label{vn}
\end{equation}
First order perturbation theory yields a
linear relation between the $c_n$'s and the $v_n$'s,
\begin{eqnarray}
c_n &=& \quad\quad
{{v_n} \over {n \omega_{eff}}} \quad\quad\quad n \quad {\rm even} \nonumber \\
c_n &=& {{\sqrt{n+1}v_{n+1} + \sqrt{n} v_{n-1}} \over
         {(n-1) \omega_{eff} }} \quad n \quad {\rm odd}
\label{cntovn}
\end{eqnarray}
Eqs. \ref{recrel} and \ref{cntovn} may be combined to form a set of
recursion relations for $v_n$.  Any choice of $v_2$ will 
generate an entire set of $v_n$'s, whose magnitude
in general grows rapidly with $n$.  
The resulting $\delta V(z)$ is ill-defined.  However, for
a given $\beta$ there is a {\it unique} choice of $v_2$ for which
$|v_n|$ uniformly decreases with increasing even $n$. (Note 
$v_n$ vanishes for odd values of $n$ and
$v_{2m}$ alternates
in sign.)  This choice may be found as follows: select
a large value of $N$ and set
$v_n =0$ for $n > N$.  The recursion relations for
$v_n$ $(0<n \le N)$ then may be written as a matrix equation
which may be solved numerically with little difficulty.
Fig. \ref{fig_vn} illustrates $v_n$ for 
$\beta=0.1$ and several increasing values of $N$.  As may be seen, 
the $v_n$ 
converge to a {\it unique} sequence as $N \rightarrow \infty$,
and the limiting values vanish rapidly with increasing $n$.

The potential we have found is the {\it unique} single particle
potential that produces the density of the harmonic atom model,
to first order in perturbation theory,
as required by the Hohenberg-Kohn theorem \cite{kohn}.  It involves
all even powers of $z$, or, equivalently, all spherical harmonics
$Y_{2l}^0$.  This demonstrates that, to reproduce the physical
density, involving $l=0$ ane $l=2$ only, requires, in the absence
of interactions, an external potential involving all even $l$-values.

\section{Perturbative Calculation of Densities and Energies for Constrained Energy
Functional}
\label{calculation}

In this Appendix, we discuss a concrete example of the ideas
developed in Section III,
providing details of how, for the helium $P$ state, our
perturbative  
Hartree approximation is modified by the introduction
of the constraint. 

We begin with the single particle Schroedinger equation,
Eq. \ref{schro}, with single particle potential
\begin{equation}
v_s(\vec{r})={{-Ze^2} \over r} +
\lambda e^2 \int d^3 r' {{n(\vec{r}')} \over {|\vec{r}-\vec{r}'|}}
 + \lambda V_R(\vec{r})
\label{vshar}
\end{equation}
To
lowest non-trivial order in $\lambda$, the corrections
to the wavefunctions $\phi_i^{(1)}$ obey Eq. \ref{wfeqn}.
The form of $U^{(1)}$ must be modified to include the
restoring potential,
$$
U^{(1)}(\vec{r}) \rightarrow U^{(1)}_R(\vec{r})
\equiv 
e^2 \int d^3 r' {{n^{(0)}(\vec{r}')} \over {|\vec{r}-\vec{r}'|}}
 + V_R(\vec{r}).
$$
Note that to this order in perturbation theory, the $l=0$ and
$l=2$ components in a spherical harmonic expansion of 
$U^{(1)}_R(\vec{r})$ come only from the Hartree potential,
and so are identical to the ones encountered in Section \ref{pert_har}.
The higher $l$ components of $U^{(1)}_R$ come only from
$V_R$.
$U^{(1)}_R(\vec{r})$ thus can be expanded as
$$
U^{(1)}_R(\vec{r}) = \Phi_0^{(1)}(r) Y_0^0(\vec{\Omega})
+ \Phi_2^{(1)}(r) Y_2^0(\vec{\Omega})
$$
$$ \quad\quad
+ v_4(r)Y_4^0(\vec{\Omega}) +v_6(r)Y_6^0(\vec{\Omega}) +\dots~.
$$
The radial functions $\Phi_n^{(1)}$ are determined fully
by the noninteracting electron density, and 
the functions $v_n(r)$ come from $V_R$.

Because all even values of $l$ appear in the decomposition of
$U^{(1)}_R$, the corrections to the wavefunctions contain
all spherical harmonics:
\begin{eqnarray}
\phi_0^{(1)}&=&y_0(r)Y_0^0(\vec{\Omega})+y_2(r)Y_2^0(\vec{\Omega})
+y_4(r)Y_4^0(\vec{\Omega}) + ...
  \nonumber \\
\phi_1^{(1)}&=&y_1(r)Y_1^0(\vec{\Omega})+y_3(r)Y_3^0(\vec{\Omega})
+y_5(r)Y_5^0(\vec{\Omega}) + \dots~.
\label{yc}
\end{eqnarray}
Noting that the radial functions $y_i(r)$ fall off very
rapidly with $l$, we retain only $l \le 4$ in
the calculations that follow.  The radial functions
$y_l$ obey Eq. \ref{y2}, with
\begin{eqnarray}
f_0(r) &=& [\varepsilon^{(1)}_0 - c^0_{00} \Phi_0^{(1)}(r)] R_{10}(r) \nonumber \\
f_1(r) &=& [\varepsilon^{(1)}_1 - c^1_{01} \Phi_0^{(1)}(r)
-c^1_{21} \Phi_2^{(1)}(r)] R_{21}(r) \nonumber \\ 
f_2(r) &=&  - c^2_{20} \Phi_2^{(1)}(r) R_{10}(r) \nonumber \\
f_3(r) &=&  - c^3_{21} \Phi_2^{(1)}(r) R_{21}(r) 
            - c^3_{41} v_4(r) R_{21}(r) \nonumber \\
f_4(r) &=&  - c^4_{40} v_4(r) R_{10}(r).
\label{f0-4}
\end{eqnarray}          
Only $f_3$ and $f_4$ are modified by $V_R$;  it follows that
the $y_0,~y_1,$ and $y_2$ are identical to the results of the Hartree
approximation without the constraint.  The equations to be solved
are closed by requiring the highest spherical harmonic component
retained, $l=4$, to vanish to first order in $\lambda$ in the density:
\begin{equation}
n_4^{(1)}(r) = c_{04}^4R_{10}(r)y_4(r) + c_{31}^4R_{21}(r)y_3(r) =0.
\label{n4}
\end{equation} 
The constrained Hartree state is found by solving Eqs. \ref{y2}, \ref{f0-4},
and \ref{n4}.  The radial functions $y_n(r)$ resulting
are presented in Fig. 2. We present
in Table 1 the energies obtained by the Hartree approximation,
the spherically-averaged Hartree approximation, and the constrained
Hartree approximation.  As can be seen, the differences among the 
three approaches only arise at order $\lambda^2$\cite{com2},
and are quite small.  Nevertheless, the constrained
Hartree approximation yields an energy considerably closer to
the unconstrained Hartree approximation than
the spherically averaged one, indicating that imposing
the symmetry by spherical averaging raises the energy
considerably more than necessary.

Finally, for comparison we also present in Table 1 analogous energies
for the results of a constrained, perturbative LDA calculation.
The method for computing the wavefunctions 
is, {\it mutatis mutandis}, 
the same as for the constrained Hartree approximation.  
As in that approximation, the
introduction of the constraint introduces little change
in the energy.

\begin{table}
\caption{Zeroth ($E^{(0)}$), first ($\lambda E^{(1)}$), and second 
($\lambda^2 E^{(2)}$) order contributions to the
He 2$P$ state energy calculated  
in perturbation theory in the electron-electron interaction
strength $\lambda$ by various methods.  
$H$=Hartree approximation,
$H-Sph$= spherically-averaged Hartree approximation,
$H-C$ = constrained Hartree approximation, $LDA$ = local density
approximation, $LDA-C$= constrained local density approximation.
Energy units are $e^2/a_B$ with $a_B=\hbar^2/me^2$ the hydrogenic Bohr radius.}
\begin{tabular}{|c|ccccc|}
~ & $H$ & $H-Sph$ & $H-C$ & $LDA$ & $LDA-C$ \\ \hline
$E^{(0)}$ & -2.5000 & -2.5000 & -2.5000 & -2.5000 & -2.5000 \\
$E^{(1)}$ &  1.3063 &  1.3063 &  1.3063 &  0.5439 &  0.5439 \\
$E^{(2)}$ & -0.4059 & -0.4039 & -0.4052 & -0.1140 & -0.1139
\end{tabular}
\end{table}

\begin{figure}

 \vbox to 6.0cm {\vss\hbox to 8cm
 {\hss\
   {\includegraphics{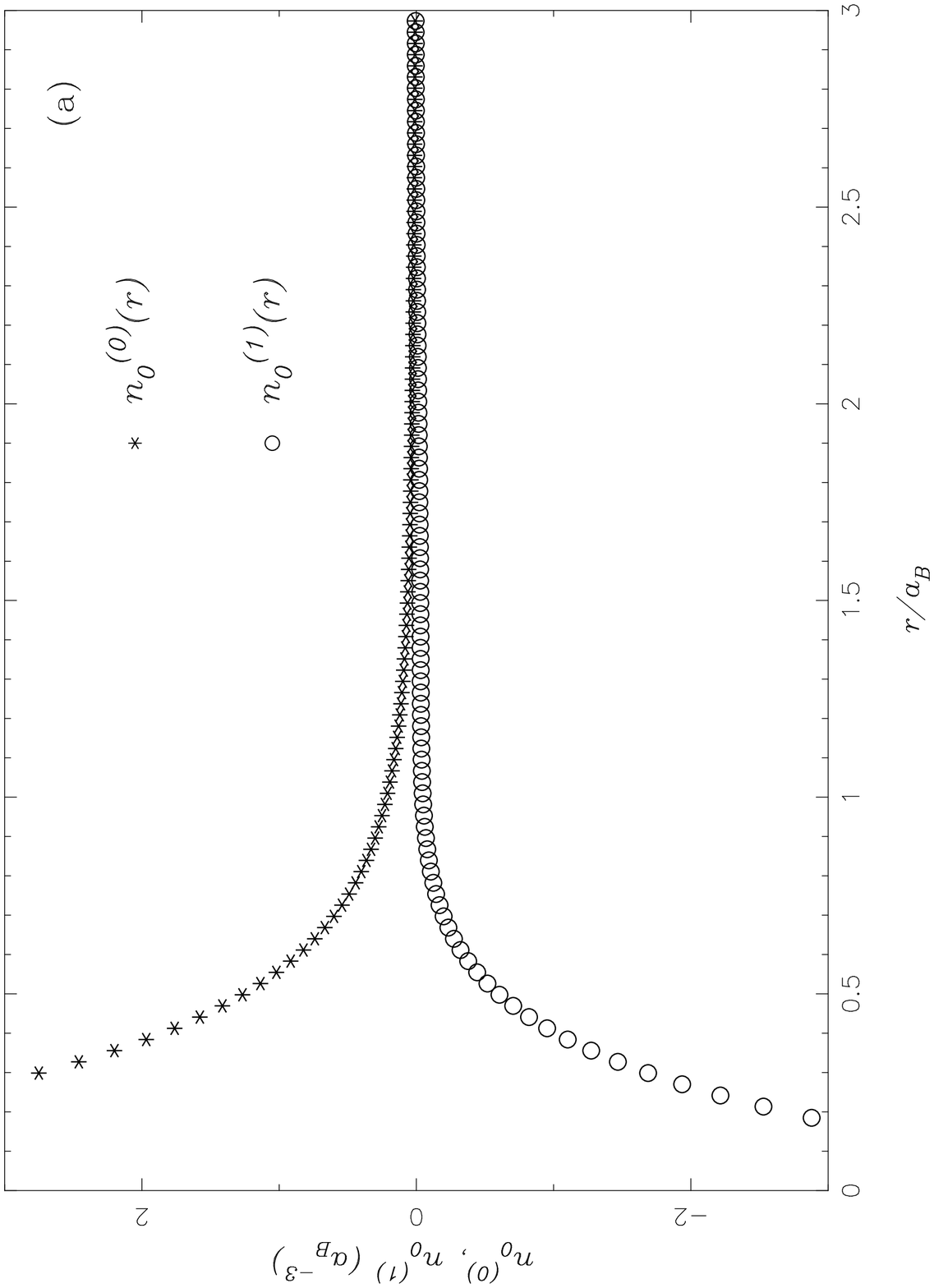}
   }
  \hss}
 }

 \vbox to 6.0cm {\vss\hbox to 8cm
 {\hss\
   {\includegraphics{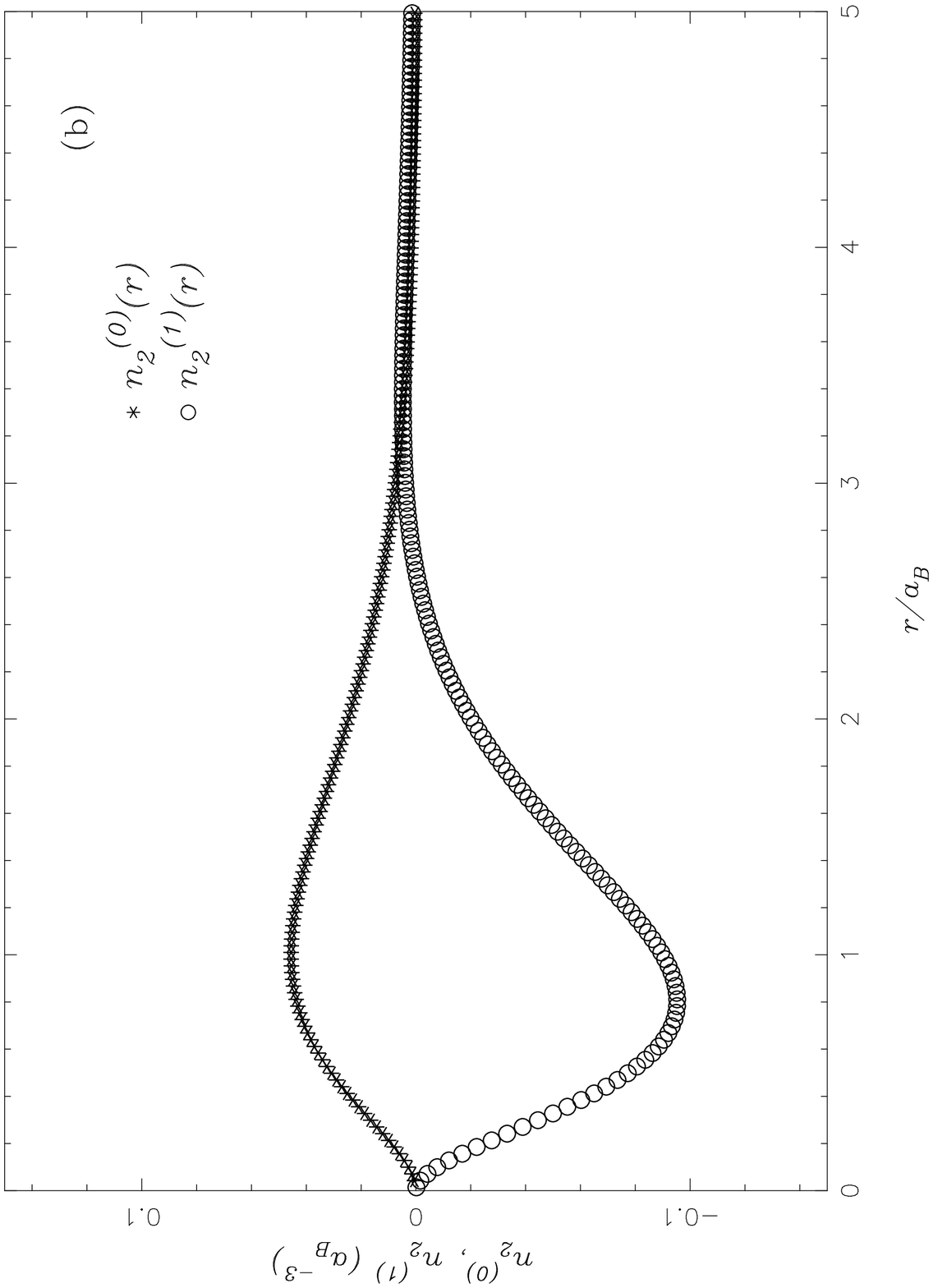}
   }
  \hss}
 }
 
 \vbox to 6.0cm {\vss\hbox to 8cm
 {\hss\
   {\includegraphics{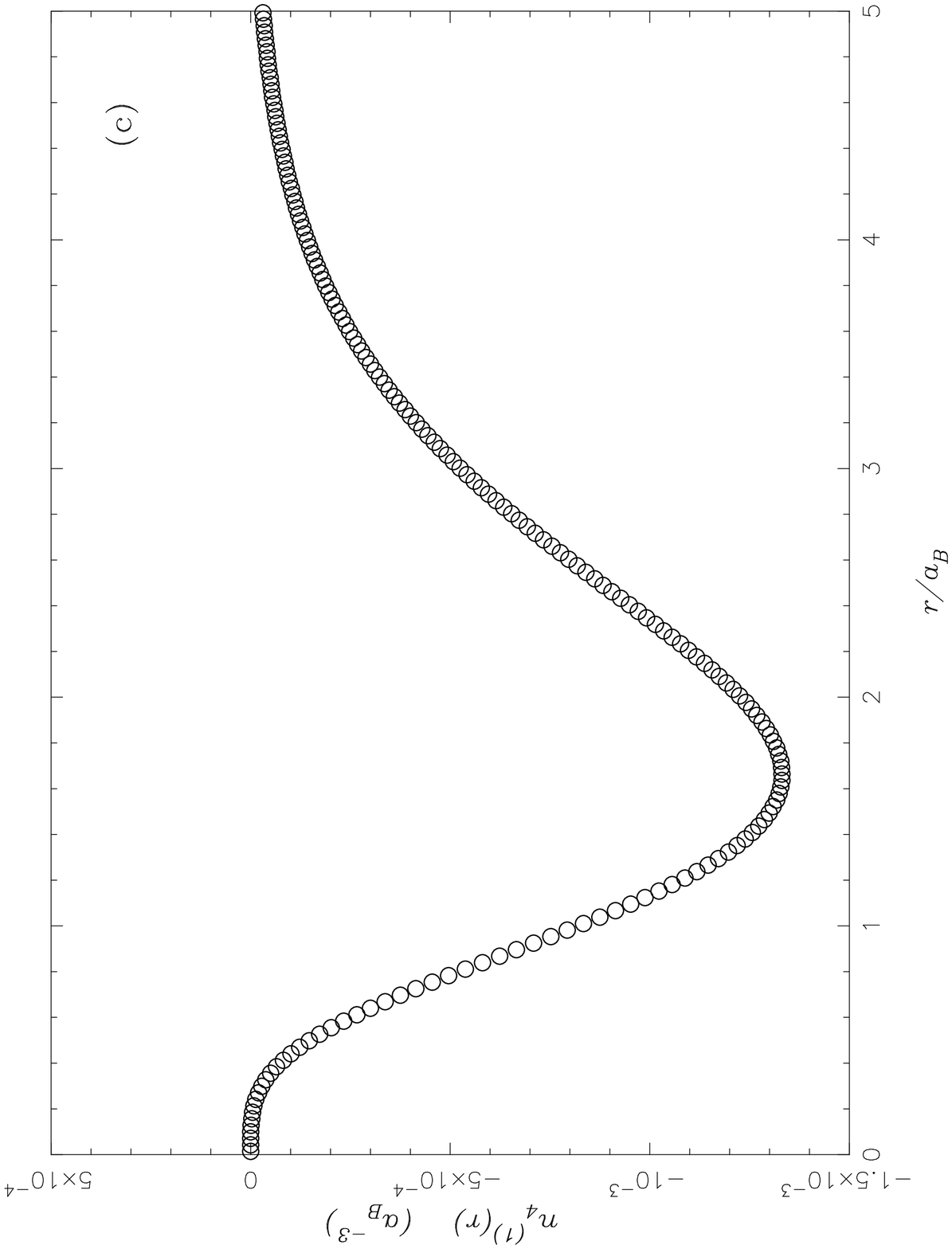}
   }
  \hss}
 }
\label{fig1}
\caption{Radial functions of the spherical harmonic decomposition  of
the density in the Hartree approximation in
a helium $P$ state.  Zeroth and first order corrections 
(designated by superscripts) in the
electron-electron interaction are illustrated.  Absolute magnitudes 
fall quickly with increasing spherical harmonic
index $l$ (designated by subscripts).  (a) $n_0^{(0)}(r),~ n_0^{(1)}(r)$; 
(b)$n_2^{(0)}(r),~ n_2^{(1)}(r)$; (c) $n_4^{(1)}(r)$.}
\end{figure}

\begin{figure}
\label{fig4}

 \vbox to 6.0cm {\vss\hbox to 8cm
 {\hss\
   {\includegraphics{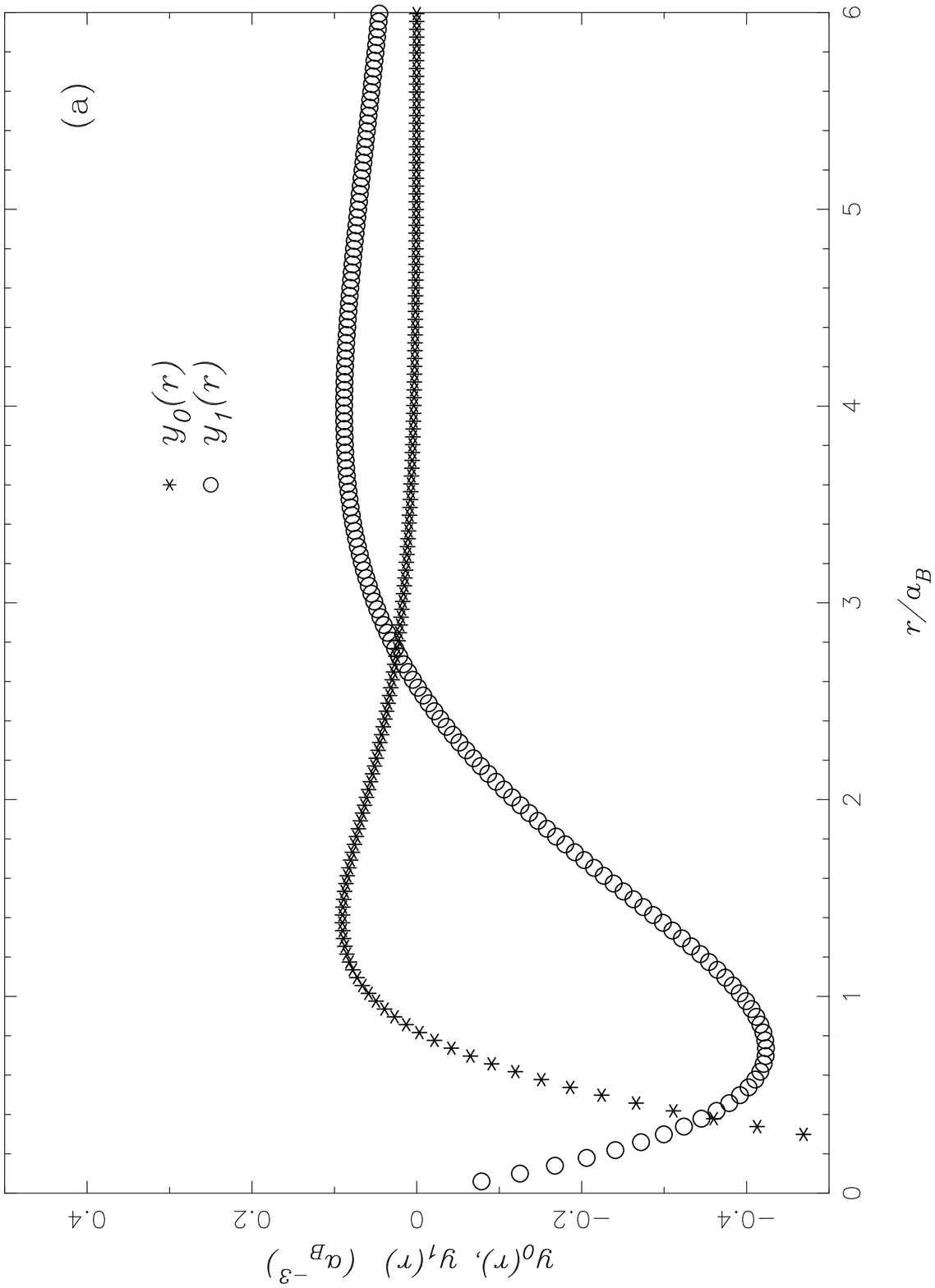}
   }
  \hss}
 } \vbox to 6.0cm {\vss\hbox to 8cm
 {\hss\
   {\includegraphics{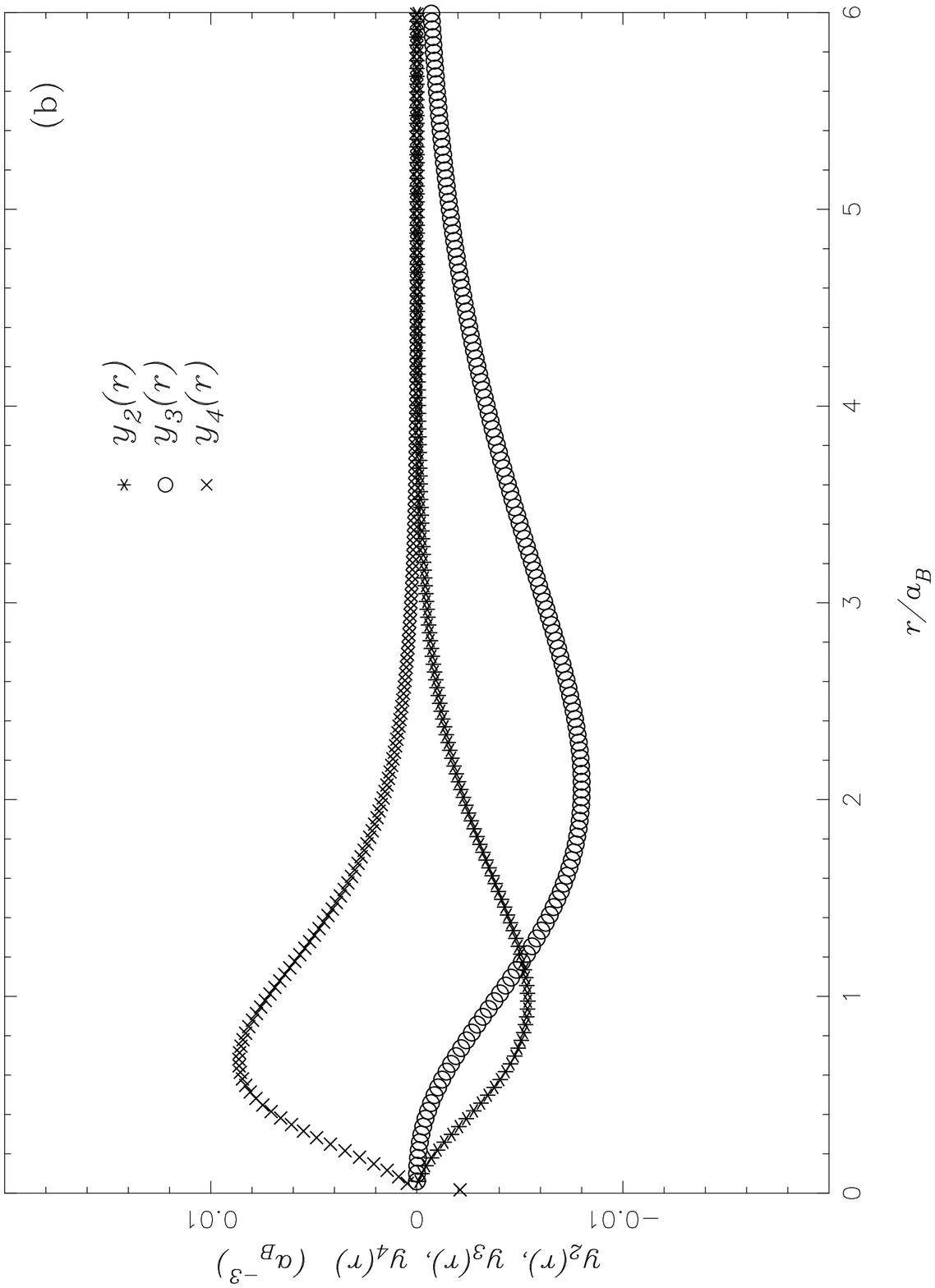}
   }
  \hss}
 }

\caption{Radial functions $y_l(r)$ produced by the constrained
Hartree approximation to first order in perturbation theory
for a helium triplet $P$ state (see text).  
(a) $y_0(r),~y_a(r)$; (b) $y_2(r),~y_3(r),~y_4(r)$. Note that the 
$l=0,1,2$ contributions are identical to the results of the standard
Hartree approximation; $l=3,4$ results are modified by the constraint.}
\end{figure}

\begin{figure}
\label{fig_vn}

 \vbox to 6.0cm {\vss\hbox to 8cm
 {\hss\
   {\includegraphics{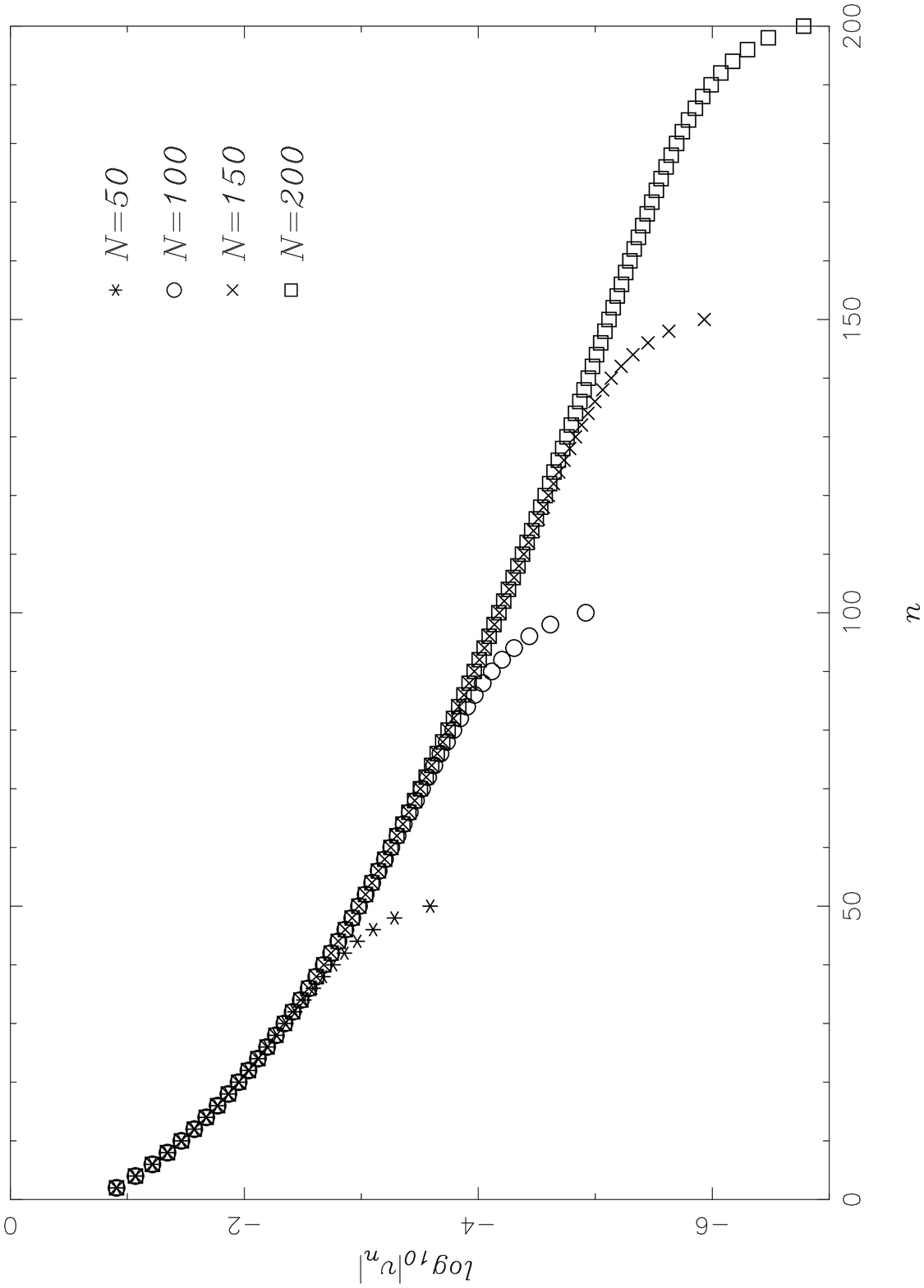}
   }
  \hss}
 }

\caption{Approximate coefficients $v_n$ in expansion of $\delta V$ given by
Eq. \ref{vn} for  $\beta=0.1$ (see text).  $v_n$ is assumed to vanish
for $n > N$ with different values of $N$ given in the figure.  The
expansion coefficients converge to a unique set of values as 
$N \rightarrow \infty$.}
\end{figure}


\begin{references}
\bibitem{com2} Second order corrections to the energy may be shown
to be derivable directly from first order corrections to the 
wavefunctions.
\end{references}
\end{document}